\documentclass{article}
\usepackage{spconf,graphicx,hyperref,xcolor}
\usepackage{epsfig, amsthm,amsmath,amssymb,epsf,cite,scalefnt,epstopdf,setspace}

\usepackage{booktabs} 
\usepackage{amsthm}
\usepackage{bbm}
\usepackage{algorithm}
\usepackage{algpseudocode}

\usepackage{graphicx}
\usepackage{subfig}
\usepackage{color}
\usepackage{url}
\usepackage{multicol}
\usepackage{svg}
\usepackage{mathtools}
\usepackage{graphicx}

\definecolor{mygray}{gray}{0.6}

\def\ba{{\mathbf{a}}}

    \def\by{{\mathbf{y}}}

    \def\bE{{\mathbf{E}}}
    
   \def\bN{{\mathbf{N}}} 
  \def\bR{{\mathbf{R}}}  
  \def\bW{{\mathbf{W}}}  \def\bY{{\mathbf{Y}}}

     \def\d4{\!\!\!\!}

  \def\-{\! - \!}  \def\+{\! + \!}  \def\={\! = \!}  \def\>{\! > \!}

\def \log{\mathrm{log}}

\newcommand{\bef}{\begin{figure}}
\newcommand{\eef}{\end{figure}}
\newcommand{\beq}{\begin{eqnarray}}
\newcommand{\eeq}{\end{eqnarray}}

\title{
Joint single-shot ToA and DoA estimation for VAA-based BLE ranging with phase ambiguity: A deep learning-based approach
}
\name{Jincheng Xie, Yili Deng, Jiguang He, Pengyu Wang, Miaomiao Dong, Rui Tang, Zhongyi Huang
\thanks{
(The first two authors contributed equally to this work. 
Corresponding author: Zhongyi Huang.)
J. Xie, Y. Deng, P. Wang and Z. Huang are with the Department of Mathematical Sciences, Tsinghua  University, Beijing, China (emails: \{xiejc22, dengyl21, py-wang24\}@mails.tsinghua.edu.cn and zhongyih@tsinghua.edu.cn).
J. He is with the School of Computing and Information Technology, Great Bay University, China (email: jiguang.he@gbu.edu.cn).
M. Dong is with the Theory Lab, Central Research Institute, 2012 Labs, Huawei Technologies Co. Ltd., Hong Kong, China (emails: dong.828599@huawei.com).
Rui Tang is with the School of Electronic Information Engineering, the Internet of Things Perception and Big Data Analysis Key Laboratory of Nanchong City, and the Key Laboratory of Electronic Information Process and Technology Application, China West Normal University, Nanchong, Sichuan 637002, China (e-mail: tangrui@cwnu.edu.cn).
The work of Rui Tang was supported in part by the National Natural Science Foundation of China under Grant 62301450.
The work of Zhongyi Huang was partially supported by the NSFC Project No. 12025104.
This work of Jiguang He was supported by Guangdong Research Team for Communication and Sensing Integrated with Intelligent Computing (Project No. 2024KCXTD047).}
}
\address{}
\vspace{-3cm}
\begin{document}
%
\maketitle
\begin{abstract}

Conventional direction-of-arrival (DoA) estimation methods rely on multi-antenna arrays, which are costly to implement on size-constrained Bluetooth Low Energy (BLE) devices.
Virtual antenna array (VAA) techniques enable DoA estimation with a single antenna, making angle estimation feasible on such devices. However, BLE only provides a single-shot two-way channel frequency response (CFR) with a binary phase ambiguity issue, which hinders the direct application of VAA.
To address this challenge, we propose a unified model that combines VAA with BLE two-way CFR, and introduce a neural network–based phase recovery framework that employs row/column predictors with a voting mechanism to resolve the ambiguity.
The recovered one-way CFR then enables super-resolution algorithms such as MUSIC for joint time-of-arrival (ToA) and DoA estimation.
Simulation results demonstrate that the proposed method achieves superior performance under non-uniform VAAs, with mean square errors approaching the Cramér–Rao bound at SNR $\geq$ 5 dB.
\end{abstract}
\begin{keywords}
Direction of arrival, time of arrival, virtual antenna array, phase recovery, BLE ranging
\end{keywords}
\vspace{-0.5cm}
\section{Introduction}
\label{sec:intro}


Accurate direction-of-arrival (DoA) and time-of-arrival (ToA) estimation underpins wireless localization and sensing~\cite{deng2025simplified}. 
In low-power systems such as Bluetooth Low Energy (BLE), phase-based ranging exploits multi-carrier phase difference (MCPD) by constructing a two-way channel frequency response (CFR) that cancels local oscillator (LO) and synchronization offsets via a Hadamard product~\cite{8885791}. 
Meanwhile, virtual antenna array (VAA) techniques synthesize spatial diversity by stitching single-antenna measurements collected along a receiver trajectory~\cite{vaa_mdpi_2020,7881965,8067441}. 
These approaches provide cost-efficient alternatives to classical super-resolution methods\cite{9838395,tang_offgrid_ANM_2013,schmidt_MUSIC_1986} (e.g., MUSIC\cite{schmidt_MUSIC_1986}) on physical multi-antenna arrays, which achieve high fidelity but remain impractical for resource-constrained devices. 
However, BLE two-way CFR aggravates multipath and reduces signal-to-noise ratio (SNR).
In the meantime recovering a one-way CFR by element-wise square root introduces a binary $\{\pm1\}$ phase ambiguity that degrades ToA and DoA estimation in practice~\cite{BoerVTC20}.
Furthermore, the irregular geometry of VAAs complicates array signal processing compared to uniform linear arrays.


Prior works explored observability and LO-offset compensation with real-world experiments~\cite{ChengTIM21}, super-resolution recovery on VAA via Jacobi--Anger expansion~\cite{WangArxiv19}, and joint ToA and DoA estimation with carrier frequency offsets compensation using large-scale arrays~\cite{GongTSP21}. 
However, these approaches often assumed multiple snapshots or uniform array structures and did not address BLE’s single-shot measurements and binary phase ambiguity. 
On the other hand, BLE MCPD ranging constructs a two-way CFR to cancel LO offsets without synchronization~\cite{8885791}, and its extension includes~\cite{BoerVTC20}, reduced-complexity estimators~\cite{ShoudhaAccess22}, and three-stage pipelines with phase unwrappingp, fade detection as well as phase errors correction ~\cite{HelwaAccess23}. 
System-level MCPD localization has also been validated on commodity BLE hardware~\cite{WC-CP2024}. Yet, these BLE methods typically rely on heuristic restoration from two-way CFR and overlook the explicit element-wise $\{\pm1\}$ ambiguity caused by square-root recovery. 
Motivated by the above drawback, in this paper we integrate VAA modeling with BLE two-way CFR to enable joint ToA and DoA estimation from single-shot measurements and introduce a learning-assisted neural voting strategy to resolve phase ambiguity. 

This paper integrates VAA modeling with BLE two-way CFR to enable joint ToA and DoA estimation from single-shot measurements. 
The square root of the two-way CFR yields a one-way CFR with binary ambiguity, whose recovery is essential before applying MUSIC. 
Directly training a single network to resolve this ambiguity performs poorly under non-uniform geometries due to irregular sampling. 
We thus decompose the task into two parts: a row-wise network predicts signs along the spatial dimension and a column-wise network along the frequency dimension. 
Their outputs are fused via majority voting to robustly estimate binary ambiguity, enabling reliable one-way CFR recovery and subsequent high-resolution ToA and DoA estimation.


\vspace{-0.5cm}
\section{System Setup}

\begin{figure}
    \centering
    \includegraphics[width=0.8\linewidth]{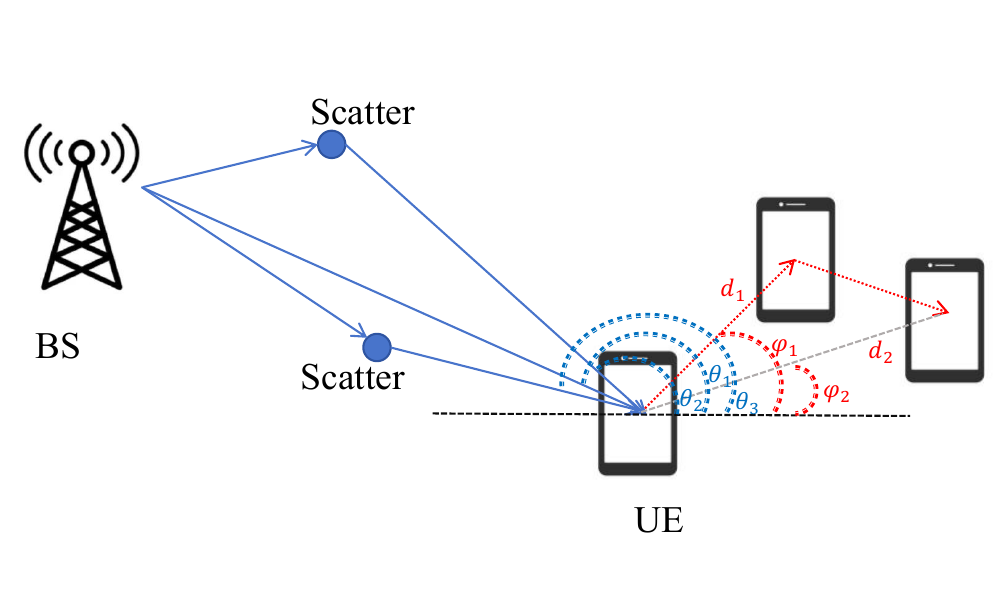}
    \vspace{-0.6cm}
    \caption{SISO multi-carrier VAA BLE communication system.}
    \label{fig:VAA}
    \vspace{-0.6cm}
\end{figure}

As shown in Fig.\ref{fig:VAA}, we consider a BLE ranging system, where the received signals are affected 
by both the geometry of the antenna array and by the LO phase offsets inherent in BLE devices. 
In this section, we first present the signal model based on a VAA 
formed by the movement of the user equipment (UE). 
We then extend this model to incorporate the BLE two-way CFR, which cancels out LO phase offsets through a Hadamard product operation.
By combining these two aspects, we obtain a unified model suitable for joint ToA and DoA estimation.

\vspace{-0.5cm}
\subsection{VAA Signal Model}
We begin with the ideal VAA representation without LO phase offsets, which captures the impact of the 
non-uniform antenna positions along the UE trajectory. 
Suppose there are $L$ multipath components. 
The signal received at the $n$-th position ($n=0,1,\dots,N-1$) 
on the $m$-th subcarrier ($m=0,1,\dots,M-1$) is
\begin{equation}
y_m^{(n)} = \sum_{\ell=1}^{L} h_{m,\ell}^{(n)} 
   e^{- j2\pi m \Delta f \tau_\ell^{(n)}} 
   + w_m^{(n)}, 
\label{eq:rx_signal}
\end{equation}
where $\tau_\ell^{(n)}$ is the ToA of the $\ell$-th path corresponding to the $n$-th position signal, $\Delta f$ is frequency spacing, and $w_m^{(n)} \sim \mathcal{CN}(0,\sigma_w^2)$ 
is additive white Gaussian noise. The channel coefficient for the $\ell$-th path is modeled as
\begin{equation}
h_{m,\ell}^{(n)} = \alpha_\ell 
   e^{j\big(\psi_\ell + \Delta \psi_{m,\ell}^{(n)}\big)},
\end{equation}
where $\alpha_\ell$ and $\psi_\ell$ are the amplitude and initial phase, 
and $\Delta \psi_{m,\ell}^{(n)}$ is the additional spatial phase shift due to the UE displacement:
\begin{equation}
\Delta \psi_{m,\ell}^{(n)} 
= \frac{2\pi d_n}{\lambda_m} \cos\!\big(\theta_\ell - \varphi_n\big),
\end{equation}
where $\lambda_m$ is the wavelength at the $m$-th subcarrier, and $\theta_\ell$ denotes the DoA of the $\ell$-th path. 
The quantities $d_n$ and $\varphi_n$ specify the displacement distance and direction of the UE at the $n$-th virtual position relative to the reference ($d_0=0, \varphi_0=0$). The sequences ${d_n}$ and ${\varphi_n}$ are measured by the UE’s IMU.
To satisfy the spatial Nyquist criterion, the distance between adjacent virtual antennas 
cannot exceed half of the carrier wavelength, i.e., $|d_{n+1} - d_n| \leq \tfrac{\lambda_\text{c}}{2}$. 
On this basis, we have
\begin{equation}
   |\Delta f \tau_l^{(n)} - \Delta f \tau_l^{(0)}| 
   \leq \frac{ \Delta f d_n }{c}  
   \leq \frac{N B}{2 M f_c} \stackrel{(a)}\approx 0, 
\label{eq: inequality}
\end{equation}
where approximation $(a)$ follows from $B \ll f_c$. 
This implies $\tau_l^{(n)} \approx \tau_l^{(0)}$, 
so we drop the superscript and write $\tau_l^{(n)} = \tau_l$. 
Furthermore, when $B \ll f_c$, $\lambda_m \approx \lambda_\text{c}$. 
The signal model in \eqref{eq:rx_signal} is thus rewritten as
\begin{equation}
    y_m^{(n)} = \sum_{\ell=1}^{L} c_\ell \,
    e^{j2\pi \left(\tfrac{d_n}{\lambda_\text{c}} \cos (\theta_\ell - \varphi_n) -  m \Delta f \tau_{\ell} \right)} 
    + w_m^{(n)}.
\end{equation}
where $c_\ell = \alpha_\ell e^{j\psi_\ell}$.
Stacking all $N$ positions and $M$ subcarriers yields the received signal matrix
\begin{equation}
\mathbf{Y} =
\sum_{\ell=1}^{L} c_\ell \,
\ba_\Theta(\theta_\ell) \,
\ba_F^H(\tau_\ell) + \mathbf{W}, 
\label{eq:vaa_matrix}
\end{equation}
the spatial steering vector is
\begin{equation}
\ba_\Theta(\theta) = 
\begin{bmatrix}
1, e^{j\frac{2\pi d_1}{\lambda_c}\cos(\theta-\varphi_1)}, \dots, e^{j\frac{2\pi d_{N-1}}{\lambda_c}\cos(\theta-\varphi_{N-1})}
\end{bmatrix}^T,
\end{equation}
the frequency steering vector is
\begin{equation}
\label{eqn:fsv}
\ba_F(\tau) = 
\begin{bmatrix}
1, e^{j2\pi \Delta f \tau}, \dots, e^{j2\pi (M-1)\Delta f \tau}
\end{bmatrix}^T,
\end{equation}
$(\cdot)^H$ and $(\cdot)^T$ denote the transpose and conjugation transpose, respectively.
Here, $\mathbf{W}$ is the additive noise matrix. 
The above model describes the VAA signal reception without considering BLE-specific impairments. 
Next, we incorporate the BLE two-way CFR construction to account for LO phase offsets.

\subsection{BLE Two-way Channel Frequency Response}
In practical BLE ranging, each one-way CFR is distorted by unknown LO phase offsets 
between the initiator and reflector. 
As these offsets vary across frequency channels, they must be canceled to obtain reliable ToA and DoA estimates. 
Let the one-way CFR from initiator to reflector $\bY^{(R)}$ be
\begin{equation}
    \bY^{(R)}= \mathbf{\Phi}\circ\bY,
\end{equation}
where $\circ$ denotes the Hadamard product, 
$\mathbf{\Phi}=[e^{j\Phi_{n,m}}]_{N\times M}$ is the LO phase offset matrix, 
and $\Phi_{n,m}$ the offset at the $(n,m)$-th element. 
Similarly, the one-way CFR from reflector to initiator $\bY^{(I)}$ is
\begin{equation}
    \bY^{(I)}= \mathbf{\Phi}^* \circ \bY,
\end{equation}
where $(\cdot)^*$ denotes element-wise conjugation. 
To mitigate LO phase offsets, we form the two-way CFR as
\begin{equation}
\begin{aligned}
\hat{\bY} & = \bY^{(R)} \circ \bY^{(I)} \\
& = \sum_{\ell_1=1}^L\sum_{\ell_2=1}^L c_{\ell_1} c_{\ell_2} 
\left( \ba_\Theta(\theta_{\ell_1}) \ba_F^H(\tau_{\ell_1})\circ 
       \ba_\Theta(\theta_{\ell_2}) \ba_F^H(\tau_{\ell_2}) \right) \\
& \quad + 2 \bW \circ \left(\sum_{\ell=1}^{L} c_\ell \,
\ba_\Theta(\theta_\ell) \,
\ba_F^H(\tau_\ell)\right) + \bW \circ \bW.
\end{aligned}
\label{two-way}
\end{equation}
This cancels the LO offsets but also increases the effective multipath order and reduces SNR. 
To address this, we extract $\tau_\ell$ and $\theta_\ell$ from the one-way CFR via
\begin{equation}
(\hat{\bY})^{1/2} = \bY \circ \bN,\quad \bN=[N_{n,m}]\in\{1,-1\}^{N\times M},
\label{one-way}
\end{equation}
where $(\cdot)^{1/2}$ denotes the element-wise square root and $\bN$ introduces a $\pm 1$ phase ambiguity. 
To eliminate this phase ambiguity, we propose an algorithm detailed below.
\vspace{-0.4cm}

\section{DOUBLE NEURAL NETWORK ALGORITHM DESIGN}
\label{sec:algorithm}
\vspace{-0.4cm}

In this section, we present the proposed algorithm for joint ToA and DoA estimation in BLE ranging with non-uniform antenna arrays. 

\vspace{-0.4cm}
\subsection{Neural Network-based Phase Recovery}
Without loss of generality, we assume the first element $N_{1,1}=1$.\footnote{Multiplying all entries by $-1$ does not affect ToA and DoA estimation.} 
To exploit structure along both dimensions, we employ two neural networks: a \textbf{row-wise network} $f_{\text{row}}(\cdot)$ and a \textbf{column-wise network} $f_{\text{col}}(\cdot)$. 
Both networks adopt the U-Net\cite{unet} architecture consisting of five down-sampling blocks with channel sizes $\{64,128,256,512,1024\}$, followed by symmetric up-sampling blocks with skip connections. 
The detailed architecture of U-Net used in this paper follows ~\cite{unet}.
On top of the final convolutional layer\cite{10.5555}, we attach a fully connected layer and apply a sigmoid activation\cite{Cybenko1989ApproximationBS}, which maps the outputs into the interval $(0,1)$ to represent probabilities.
The input to each network is the phase tones along one dimension (across subcarriers or across antennas), and the output is the probability that each element in the corresponding row or column of $\bN$ has the same sign as its first entry.
The networks output probabilities for $\{-1,+1\}$ per row/column. 
We train the networks using supervised learning with cross-entropy loss:
\begin{equation}
\begin{aligned}
    \mathcal{L}_{\text{row}} 
    &= -\sum_{n,m}^{N,M} \Big( 
        q_{n,m}\log \big(f_\text{row}(\hat{\bY}_{[n,:]})_m\big) \\
    &\qquad\qquad + (1 - q_{n,m}) \log \big(1 - f_\text{row}(\hat{\bY}_{[n,:]})_m\big)
    \Big).
\end{aligned}
\end{equation}
\begin{equation}
\begin{aligned}
    \mathcal{L}_{\text{col}} 
    &= -\sum_{n,m}^{N,M} \Big( 
        p_{n,m}\log \big(f_\text{col}(\hat{\bY}_{[:,m]})_n\big) \\
    &\qquad\qquad + (1 - p_{n,m}) \log \big(1 - f_\text{col}(\hat{\bY}_{[:,m]})_n\big)
    \Big).
\end{aligned}
\end{equation}
where 
\vspace{-0.5cm}
\begin{equation}
\begin{aligned}
q_{n,m} &=
\begin{cases}
1, & \text{if } \dfrac{N_{n,m}}{N_{n,1}} = 1, \\
0, & \text{otherwise},
\end{cases}
\!\quad\!
p_{n,m} &=
\begin{cases}
1, & \text{if } \dfrac{N_{n,m}}{N_{1,m}} = 1, \\
0, & \text{otherwise}.
\end{cases}
\end{aligned}
\end{equation}
is the ground-truth label,
 $\hat{\bY}_{[n,:]}$ and $\hat{\bY}_{[:,m]}$ denote the $n$-th row and 
the $m$-th column of $\hat{\bY}$, respectively, while 
$f_{\text{row}}(\cdot)_m$ and $f_{\text{col}}(\cdot)_n$ denote the $m$-th and $n$-th output elements of the corresponding neural networks.

\subsection{Voting Mechanism}
In this stage we recover the ambiguity matrix $\bN$ using the predictions of
the row- and column-wise networks. For the row-wise predictor, when
$f_{\text{row}}(\hat{\bY}_{[n,:]})_m>0.5$, we regard $N_{n,m}$ as having the
same sign as $N_{n,1}$, which can be expressed as
\begin{equation}
N_{n,m} = 
g\left(f_{\text{row}}(\hat{\bY}_{[n,:]})_m\right)\,N_{n,1},
\end{equation}
where $g(x) = \text{sign}(x -0.5)$.
Similarly, for the column-wise predictor, when
$f_{\text{col}}(\hat{\bY}_{[:,m]})_n>0.5$, we interpret $N_{n,m}$ as having
the same sign as $N_{1,m}$.

Because the subcarrier spacing is fixed while the VAA trajectory introduces
additional randomness, the row-wise predictor is empirically more reliable. 
Therefore, we enforce that the recovered $\bN$ must be consistent with the
row-wise outputs. This reduces the problem to determining only the signs of
the first element of each row. Since we assume $N_{1,1}=1$, the first row
is obtained as $N_{[1,:]} = g\left(f_{\text{row}}(\hat{\bY}_{[1,:]})\right).$
For rows $n\geq 2$, the sign of the first element $N_{n,1}$ can be inferred
by combining both predictors. Specifically, from each column $m$ we obtain a
candidate estimate
\begin{equation}
\small
\bar{N}_{n,1,m} =
g\!\left(f_{\text{row}}(\hat{\bY}_{[1,:]})_m\right)\,
g\!\left(f_{\text{col}}(\hat{\bY}_{[:,m]})_n\right)\,
g\!\left(f_{\text{row}}(\hat{\bY}_{[n,:]})_m\right),
\label{m cand}
\end{equation}
where the first term returns $N_{1,1}N_{1,m}$, 
the second term returns $N_{1,m}N_{n,m}$, and
the third term returns $N_{n,m}N_{n,1}$.
These terms check whether the elements in $\bN$ share the same sign.
Since $f_{\text{row}} (\cdot)$ is empirically more reliable, we enforce its predictions with higher precision compared to those of $f_{\text{col}} (\cdot)$.
Thus, determining the sign of the first element $N_{n,1}$ suffices to recover the entire $n$-th row according to $f_{\text{row}}(\cdot)$. 
Finally, we aggregate these $M$ candidates by majority voting:
\begin{equation}
N_{n,1} = \text{sign}\!\left(\sum_{m=1}^M \bar{N}_{n,1,m} \right).
\end{equation}
This voting strategy yields a robust estimate of $N_{n,1}$ and hence allows us to reconstruct the entire ambiguity matrix $\bN$ consistently.
We predict other elements in $\bN$ by
\begin{equation}
\label{eqn:recover N}
    N_{n,m} = g(f_{\text{row}}(\hat{\bY}_{[n,:]})_m)N_{n,1}, \forall n>1,m>1.
\end{equation}

\subsection{Parameter Estimation via MUSIC}
After recovering $\bN$, we obtain the corrected one-way CFR:
\begin{equation}
    \widetilde{\bY} = (\hat{\bY})^{1/2} \circ \bN.
\end{equation}
We then apply a two-dimensional MUSIC algorithm to estimate the set of ToAs and DoAs. 
Specifically, we form the covariance matrix
\begin{equation}
    \bR = \frac{1}{M} \sum_{m=0}^{M-1} \widetilde{\by}_{[:,m]} \widetilde{\by}_{[:,m]}^H,
\end{equation}
where $\widetilde{\by}_{[:,m]}$ is the $m$-th column of $\widetilde{\bY}$. 
The MUSIC pseudospectrum is computed as
\begin{equation}
    P(\theta,\tau) = \frac{1}{\ba_\Theta^H(\theta)\, \bE_n \bE_n^H \,\ba_F(\tau)},
\end{equation}
where $\bE_n$ is the noise subspace basis obtained from eigen-decomposition of $\bR$. 
The estimated parameters $\{\hat{\theta}_\ell, \hat{\tau}_\ell\}$ correspond to the peaks of $P(\theta,\tau)$.







\section{SIMULATION STUDIES}
\label{sec:simulation}





We consider a SISO multi-carrier system in a $40 \,\text{m} \times 40 \,\text{m}$ 2D scenario, where the BS at the origin communicates with a UE in the far field. 
The channel includes one LoS path and $1$--$3$ NLoS paths generated by randomly placed scatterers, with the UE--BS distance uniformly distributed in $[20,30]$~m. 
A VAA is formed by single-snapshot signals at $N=16$ positions along the UE’s trajectory, where the inter-element spacing is uniformly drawn from $[\lambda/4,\lambda/2]$ with $\lambda=0.125$~m, and the moving direction is uniformly distributed within $[-\pi/4,\pi/4]$. 
The system employs $M=80$ subcarriers with frequency spacing $\Delta f=1$~MHz, and the received SNR is defined as $\mathrm{SNR}=10\log_{10}(1/\sigma_w^2)$. These settings are used to evaluate the proposed ToA and DoA estimators under multipath and noise.
We generated a total of 8000 VAA channel realizations, of which 60\% were used for training, 20\% for validation, and the remaining 20\% for testing.

\begin{figure}
    \centering
    \includegraphics[width=0.9\linewidth]{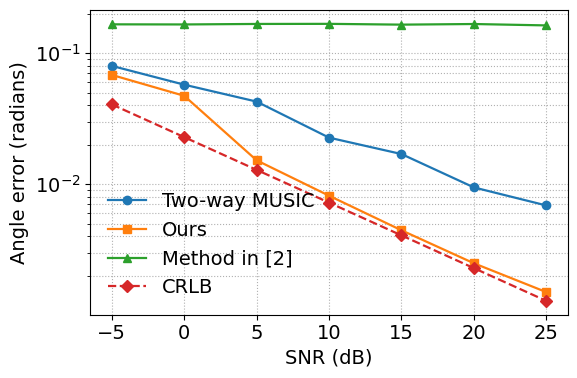}
    \vspace{-0.5cm}
    \caption{MSE plot of DoA estimation under different SNRs.}
    \label{fig:deg mse}
\end{figure}

\begin{figure}
    \centering
    \includegraphics[width=0.9\linewidth]{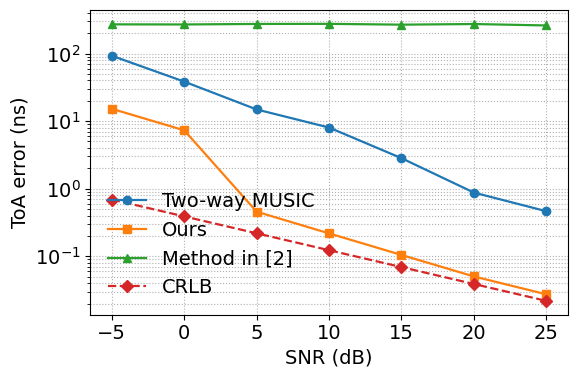}
    \vspace{-0.5cm}
    \caption{MSE plot of ToA estimation under different SNRs.}
    \label{fig:toa mse}
\end{figure}

We compare the proposed method against three baselines: 
i) the two-way CFR MUSIC algorithm, 
ii) the phase-continuity based approach for ambiguity resolution in uniform arrays from~\cite{8885791}, 
and iii) the theoretical Cramér–Rao lower bound (CRLB) computed using the method in~\cite{10622274}. 
Figs.~\ref{fig:deg mse} and~\ref{fig:toa mse} present the mean squared error (MSE) performance for angle and delay estimation, respectively, under different SNRs. 
As shown in Fig.~\ref{fig:deg mse}, our method significantly outperforms two-way MUSIC and the continuity-based approach across all SNRs. 
The performance of the proposed estimator closely approaches the CRLB, demonstrating its statistical efficiency. 
Similarly, Fig.~\ref{fig:toa mse} shows that the proposed method achieves much lower ToA estimation error compared to the baselines, especially in low-to-moderate SNR regimes. 
From the figures, it can be observed that the continuity-based method in~\cite{8885791} completely fails under non-uniform antenna array geometries, leading to large estimation errors. In contrast, our proposed approach consistently achieves lower MSE than the two-way CFR MUSIC algorithm across all SNRs. This demonstrates that the neural voting framework effectively resolves the binary phase ambiguity, enables reliable recovery of the one-way CFR, and achieves accurate super-resolution ToA and DoA estimation.

\section{Conclusion}
\label{sec:Conclusion}


This paper presented a joint ToA and DoA estimation method for BLE ranging by integrating VAA modeling with two-way CFR. 
A unified model was established, where the square-root operation introduces an element-wise $\{\pm1\}$ ambiguity. 
To address this, we proposed a neural voting framework with row- and column-wise predictors, whose fusion enables robust ambiguity recovery. 
The corrected one-way CFR supports super-resolution estimation, and simulations show that the method achieves high accuracy under multipath and non-uniform geometries, with MSEs close to the CRLB.

\newpage

\vfill\pagebreak

\bibliographystyle{IEEEbib}
\bibliography{strings,refs}

@ARTICLE{schmidt_MUSIC_1986,
  author={Schmidt, R.},
  journal={IEEE Transactions on Antennas and Propagation}, 
  title={Multiple emitter location and signal parameter estimation}, 
  year={1986},
  volume={34},
  number={3},
  pages={276-280},
  doi={10.1109/TAP.1986.1143830}}

@Article{vaa_mdpi_2020,
AUTHOR = {Pasternak, Yuri G. and Ashikhmin, Aleksander V. and Rembovsky, Yuri A. and Fedorov, Sergey M. and Zhuravlev, Dmitry V.},
TITLE = {Virtual Antenna Array for Minimization of {DOA} Estimation Systematic Error Caused by Scattering of Incident Waves on Antenna Carrier Body},
JOURNAL = {Electronics},
VOLUME = {9},
YEAR = {2020},
NUMBER = {2},
ARTICLE-NUMBER = {308},
URL = {https://www.mdpi.com/2079-9292/9/2/308},
ISSN = {2079-9292},
DOI = {10.3390/electronics9020308}
}

@misc{tang_offgrid_ANM_2013,
      title={Compressed Sensing off the Grid}, 
      author={Gongguo Tang and Badri Narayan Bhaskar and Parikshit Shah and Benjamin Recht},
      year={2013},
      eprint={1207.6053},
      archivePrefix={arXiv},
      primaryClass={cs.IT},
      url={https://arxiv.org/abs/1207.6053}, 
}

@ARTICLE{ChengTIM21,
  author={Cheng, Jianqiao and Guan, Ke and Quitin, François},
  journal={IEEE Transactions on Instrumentation and Measurement}, 
  title={Direction-of-Arrival Estimation With Virtual Antenna Array: Observability Analysis, Local Oscillator Frequency Offset Compensation, and Experimental Results}, 
  year={2021},
  volume={70},
  pages={1-13},
  doi={10.1109/TIM.2021.3088434}}

@ARTICLE{WangArxiv19,
  author={Wang, Yue and Zhang, Yu and Tian, Zhi and Leus, Geert and Zhang, Gong},
  journal={IEEE Journal of Selected Topics in Signal Processing}, 
  title={Super-Resolution Channel Estimation for Arbitrary Arrays in Hybrid Millimeter-Wave Massive {MIMO} Systems}, 
  year={2019},
  volume={13},
  number={5},
  pages={947-960},
  doi={10.1109/JSTSP.2019.2937632}}

@ARTICLE{GongTSP21,
  author={Gong, Ziyi and Wu, Liang and Zhang, Zaichen and Dang, Jian and Zhu, Bingcheng and Jiang, Hao and Li, Geoffrey Ye},
  journal={IEEE Transactions on Signal Processing}, 
  title={Joint {TOA} and {DOA} Estimation With {CFO} Compensation Using Large-Scale Array}, 
  year={2021},
  volume={69},
  number={},
  pages={4204-4218},
  doi={10.1109/TSP.2021.3095722}}

@INPROCEEDINGS{BoerVTC20,
  author={Boer, Pepijn and Romme, Jac and Govers, Jochem and Dolmans, Guido},
  booktitle={2020 IEEE 91st Vehicular Technology Conference (VTC2020-Spring)}, 
  title={Performance of High-Accuracy Phase-Based Ranging in Multipath Environments}, 
  year={2020},
  pages={1-5},
  doi={10.1109/VTC2020-Spring48590.2020.9128721}}

@ARTICLE{ShoudhaAccess22,
  author={Shoudha, Shamman Noor and Van Marter, Jayson P. and Helwa, Sherief and Dabak, Anand G. and Torlak, Murat and Al-Dhahir, Naofal},
  journal={IEEE Access}, 
  title={Reduced-Complexity Decimeter-Level Bluetooth Ranging in Multipath Environments}, 
  year={2022},
  volume={10},
  pages={38335-38350},
  doi={10.1109/ACCESS.2022.3165653}}

@ARTICLE{HelwaAccess23,
  author={Helwa, Sherief and Van Marter, Jayson P. and Shoudha, Shamman Noor and Ben-Shachar, Matan and Alpert, Yaron and Dabak, Anand G. and Torlak, Murat and Al-Dhahir, Naofal},
  journal={IEEE Access}, 
  title={Bridging the Performance Gap Between Two-Way and One-Way {CSI}-Based 5 {GHz} {WiFi} Ranging}, 
  year={2023},
  volume={11},
  pages={70023-70039},
  doi={10.1109/ACCESS.2023.3287850}}

@article{WC-CP2024,
author = {Yan, Jinjin and Zhang, Manyu and Yang, Jinquan and Mihaylova, Lyudmila and Yuan, Weijie and Li, You},
year = {2024},
month = {10},
pages = {354},
title = {{WC-CP}: A Bluetooth Low Energy Indoor Positioning Method Based on the Weighted Centroid of the Convex Polygon},
volume = {13},
journal = {ISPRS International Journal of Geo-Information},
doi = {10.3390/ijgi13100354}
}

@INPROCEEDINGS{10622274,
  author={Deng, Yili and Luo, Baojia and Xie, Jincheng and Dong, Miaomiao and Huang, Zhongyi and Chen, Xiang and Han, Wei},
  booktitle={IEEE International Conference on Communications}, 
title={Super-Resolution Joint {DoA} and {ToA} Estimation with Virtual Antenna Array}, 
  year={2024},
  volume={},
  number={},
  pages={3140-3145},
  doi={10.1109/ICC51166.2024.10622274}}

@article{deng2025simplified,
  title={A Simplified Algorithm for Joint Real-Time Synchronization, {NLoS} Identification, and Multi-Agent Localization},
  author={Deng, Yili and Fan, Jie and He, Jiguang and Luo, Baojia and Dong, Miaomiao and Huang, Zhongyi},
  journal={IEEE Transactions on Vehicular Technology},
  year={2025},
  publisher={IEEE}
}

@INPROCEEDINGS{9838395,
  author={Luo, Baojia and Dong, Miaomiao and Wu, Hao and Li, Yue and Yang, Lu and Chen, Xiang and Bai, Bo},
  booktitle={IEEE International Conference on Communications}, 
  title={Reconfigurable Intelligent Surface Assisted Millimeter Wave Indoor Localization Systems}, 
  year={2022},
  volume={},
  number={},
  address={Seoul, Korea},
  pages={4535-4540},
  doi={10.1109/ICC45855.2022.9838395}}

@INPROCEEDINGS{7881965,
  author={Quitin, Francois and Govindaraj, Vivek and Zhong, Xionghu and Tay, Wee Peng},
  booktitle={2016 IEEE 84th Vehicular Technology Conference (VTC-Fall)}, 
  title={Virtual Multi-Antenna Array for Estimating the Angle-of-Arrival of a {RF} Transmitter}, 
  year={2016},
  volume={},
  number={},
  pages={1-5},
  doi={10.1109/VTCFall.2016.7881965}}

@ARTICLE{8067441,
  author={Quitin, François and De Doncker, Philippe and Horlin, François and Tay, Wee Peng},
  journal={IEEE Transactions on Vehicular Technology}, 
  title={Virtual Multiantenna Array for Estimating the Direction of a Transmitter: System, Bounds, and Experimental Results}, 
  year={2018},
  volume={67},
  number={2},
  pages={1510-1520},
  doi={10.1109/TVT.2017.2762728}}

@INPROCEEDINGS{8885791,
  author={Zand, Pouria and Romme, Jac and Govers, Jochem and Pasveer, Frank and Dolmans, Guido},
  booktitle={2019 IEEE Wireless Communications and Networking Conference (WCNC)}, 
  title={A high-accuracy phase-based ranging solution with Bluetooth Low Energy ({BLE})}, 
  year={2019},
  volume={},
  number={},
  pages={1-8},
  doi={10.1109/WCNC.2019.8885791}}

@inproceedings{10.5555,
author = {Krizhevsky, Alex and Sutskever, Ilya and Hinton, Geoffrey E.},
title = {ImageNet classification with deep convolutional neural networks},
year = {2012},
publisher = {Curran Associates Inc.},
address = {Red Hook, NY, USA},
booktitle = {Proceedings of the 26th International Conference on Neural Information Processing Systems - Volume 1},
pages = {1097–1105},
numpages = {9},
location = {Lake Tahoe, Nevada},
series = {NIPS'12}
}

@article{Cybenko1989ApproximationBS,
  title={Approximation by superpositions of a sigmoidal function},
  author={George V. Cybenko},
  journal={Mathematics of Control, Signals and Systems},
  year={1989},
  volume={2},
  pages={303-314},
  url={https://api.semanticscholar.org/CorpusID:3958369}
}

@article{unet,
  author       = {Olaf Ronneberger and
                  Philipp Fischer and
                  Thomas Brox},
  title        = {U-{Net}: Convolutional Networks for Biomedical Image Segmentation},
  journal      = {CoRR},
  volume       = {abs/1505.04597},
  year         = {2015},
  url          = {http://arxiv.org/abs/1505.04597},
  eprinttype    = {arXiv},
  eprint       = {1505.04597},
  bibsource    = {dblp computer science bibliography, https://dblp.org}
}

\end{document}